\documentstyle[11pt]{article}

\def\a{\alpha} 
\def\b{\beta} 
 
\def\e{\epsilon}

\def\m{\mu} 
\def\n{\nu}

\def\s{\sigma} 
\def\t{\tau} 
 
 \def\O{\Omega}
\def\be{\begin{equation}}
\def\ee{\end{equation}}
     
\begin{document}

\begin{titlepage}

\begin{flushright}

BRX TH-427\\

\end{flushright}

\begin{center}
{\large\bf Hamiltonian Electric/Magnetic 
Duality and Lorentz Invariance}

\end{center}
\vspace{3cm}

\begin{center}
{\large
S. Deser and  \"{O}. Sar{\i}o\~{g}lu}
\footnote{deser,sarioglu@binah.cc.brandeis.edu}
\end{center}
\vspace{1cm}

\begin{center}{\sl
Department of Physics, Brandeis University,\\
Waltham, MA 02254, U.S.A.}
\end{center}
\vspace{3cm}

\begin{abstract}
In (3+1) Hamiltonian form, the conditions for the electric/magnetic 
invariance of generic self-interacting gauge vector actions  
and the definition of the duality generator are obvious. 
Instead, (3+1) actions are not intrinsically Lorentz invariant.  
Imposing the Dirac--Schwinger stress tensor commutator requirement 
to enforce the latter yields a differential constraint on the
Hamiltonian which translates into the usual Lagrangian form of 
the duality invariance condition obeyed by Maxwell and Born-Infeld
theories. We also discuss covariance properties of some analogous
scalar models.
\end{abstract}

\end{titlepage}

The conditions for duality invariance of $D=4$ vector gauge 
theories \cite{DT} and more generally of $n$-form models in 
$4 n$ dimensions \cite{DGHT} are well known. 
Although the duality transformation generators, $\Omega$,
are necessarily defined canonically (rather than covariantly), 
as is the verification of invariance, namely commutation 
of $\Omega$ with the Hamiltonian, the invariance criterion  
is usually stated covariantly \cite{GR,PS} as a constraint
on the Lagrangian. In the present note, we start canonically
with an {\it a priori} purely 3--invariant formulation: 
Here $\Omega$ and the invariance requirement will be easy 
to find. Instead, the hard part will be to impose Lorentz 
invariance, thereby recovering the covariant criterion. Our ingredients 
are simple: (a) the known relation \cite{DM} between Lagrangian 
and (3+1) descriptions for any vector field action depending on 
$F_{\m\n}$ (but, not, for simplicity on derivatives of 
$F_{\m\n}$), (b) the classic Dirac--Schwinger 
local stress tensor commutator criterion \cite{DS} for 
Lorentz invariance of systems of spin $\leq$ 1.

Any gauge invariant second-order
action, {\it i.e.}, one in which $F_{\mu\nu} \equiv  
\partial_{\mu} A_{\nu} - 
\partial_{\nu} A_{\mu}$ are dependent variables,
\be
I[A_{\mu}] = \int d^4x L(\alpha,\beta)  \;\; , \;\;\;\;
\alpha \equiv \frac{1}{2} F_{\mu\nu} F^{\mu\nu} \;\; , 
\;\;\;\; \beta \equiv \frac{1}{4} F_{\m\n} ~^*\!F^{\m\n} 
\;\; , \;\;\;\;
 ^*\!F^{\m\n} \equiv \frac{1}{2} \; \e^{\m\n\s\t} F_{\s\t} 
\;\; , 
\label{act}
\ee
has the equivalent first order form 
\footnote{The special cases for which $L - 
2(\alpha L_{\alpha} + \beta L_{\beta}) = 0$, such as
$L = \sqrt{a \a+ b \b} f(\frac{\a}{\b})$, where $f$ is 
arbitrary, has no Maxwell limit, but it can also be 
accommodated by use of Lagrange multipliers.} \cite{DM},
\be
\tilde{I}[F^{\mu\nu}, A_{\sigma}] = \int d^4x \tilde{L} = 
\int d^4x \left( (L_{\alpha} F^{\mu\nu} + \frac{1}{2}
L_{\beta} ~^*\!F^{\m\n})
(\partial_{\mu} A_{\nu} - \partial_{\nu} A_{\mu}) + L -
2(\alpha L_{\alpha} + \beta L_{\beta}) \right) 
\label{firact}  
\ee
where $F_{\mu\nu}$ and $A_{\sigma}$ are to be varied 
independently. Here $\tilde{L}$ depends on $A_\m$ 
{\it only} in the first term;
the $(\a,\b)$ only involve $F_{\mu\nu}$; subscripts on 
$L$ mean differentiation with respect to $\a$ or $\b$. 
The $F_{\mu\nu} = \partial_{\mu} A_{\nu} - \partial_{\nu} 
A_{\mu}$ relation emerges
as the field equation from varying $F_{\mu\nu}$.
Our conventions $\e^{0123} = +1$, $\eta = (-,+,+,+)$ 
imply $\a = {\bf B}^2 - {\bf E}^2$, $\b = -{\bf B} \cdot 
{\bf E}$ with 
$E^i \equiv F^{0i}$, $B^i \equiv \frac{1}{2} \; \e^{ijk} 
F_{jk}$.

The Gauss constraint from varying $A_0$ implies that 
the ``true" electric field variable conjugate to ${\bf A}$, 
$-{\bf D} \equiv 2 L_{\alpha} {\bf E} + L_{\beta} {\bf B}$ is 
transverse, so that we get the gauge invariant canonical action
\be
\tilde{I} = \int d^4x
\left( -{\bf D}^{T} \cdot {\bf \dot A}^{T} - \tilde{H} \right) \;\;,
\ee
\be
\tilde{H}[{\bf D}^{T}, {\bf A}^{T}] = 
- 2 {\bf B}^{2} L_{\alpha} - L + 2 \alpha L_{\alpha} +
\beta L_{\beta} \;\; .
\label{ham}
\ee
Defining a second potential ${\bf Z}$ for the transverse 
${\bf D}$ field  by ${\bf D} \equiv {\bf \nabla} \times 
{\bf Z}$ puts $({\bf D},{\bf B})$ on a symmetric footing, 
as is even clearer if one defines \cite{DGHT} the doublet 
${\bf A}^{a} \equiv ({\bf A}, {\bf Z})$,
${\bf B}^{a} \equiv ({\bf B}, {\bf D})$
\be
\tilde{I} = \int d^4x \left( \frac{1}{2} \e_{ab} \; {\bf B}^{a} 
\cdot {\bf \dot A}^{b} - \tilde{H} ({\bf B}^{a}) \right) \;\; .
\label{actab}
\ee

We have just seen how to recast a manifestly Lorentz 
invariant action (\ref{act}) or its alternate form 
(\ref{firact}) to Hamiltonian form (\ref{actab}), whose 
(highly non-manifest) invariance is guaranteed by the precise 
form of the Hamiltonian $\tilde{H}$ in (\ref{ham}). Suppose 
instead that we begin with the (3+1) Hamiltonian form (\ref{actab}) 
without any such {\it a priori} Lorentz invariance properties.
Instead, it describes a pair of 3--vectors ${\bf B}^{a}$ 
in a 3--invariant fashion, but in general does {\bf not} 
correspond to a Minkowski invariant model. [Even in the 
simplest, Maxwell case, any explicit reconstruction would 
require turning the transverse ${\bf A}^{T}$ into a 4--vector, 
recognizing that $({\bf B},{\bf D})$ are parts of a 6--tensor, etc.] 
Fortunately, there is a direct Lorentz invariance criterion, 
for spin $\leq$ 1 systems, that requires knowledge only of the 
energy and momentum densities. While these quantities are not 
uniquely defined from their spatial integrals,
it suffices to find an appropriate gauge invariant set. 
Clearly $T^0_0$ can be taken to be the Hamiltonian density
$\tilde{H} ({\bf B}^{a})$. As Dirac has taught us, the 
momentum of any system is dynamics-independent: ${\bf P} = 
\int d^3x \, \sum_{i} \pi^i (-{\nabla} \phi_i)$. In our case,
we may therefore take the usual gauge invariant
choice $T^{0i} = ({\bf D} \times {\bf B})^i =
\frac{1}{2} (\e_{ba} {\bf B}^{a} \times
{\bf B}^{b})^i$ whose integral is also {\bf P}; in the
absence of gravitation, there is no unique choice for the
densities. The Dirac--Schwinger \cite{DS} Lorentz invariance 
condition,
\be
[ T^{00}({\bf r}), T^{00}({\bf r^\prime}) ] =
\left( T^{0i}({\bf r}) + T^{0i}({\bf r^\prime}) \right) 
\partial_i
\delta^3({\bf r} - {\bf r^\prime}) \;\; ,
\label{DiSc}
\ee
is to be computed through the canonical Poisson bracket (or 
commutation) relation 
$[ B^a_i({\bf r}), B^b_j({\bf r^\prime}) ] =
\e^{ba} \e_{ijk} \partial^k \delta^3({\bf r} - 
{\bf r^\prime}) $, with both sides transverse.

[It should be noted that (\ref{DiSc}) (or its half-integrated
form) is an ``on-shell" condition. Thus (\ref{DiSc}) can
verify Lorentz covariance of Hamiltonian forms even if these
do not have a simple ``off-shell" covariant equivalent. An
illustration is provided by the $D=2$ self-dual scalar
field \cite{Jac}, $I=\int d^2x (\pi {\dot \phi} - \pi \phi')$
where $T^{00}=T^{01}= \pi \phi'$ and (\ref{DiSc}) is 
manifestly obeyed. However, there is no underlying
$L((\partial_{\m} \phi)^2)$ form.]

We are now in a position to first impose the duality 
(trivial) and then the Lorentz (nontrivial) invariance 
on our system (\ref{actab}).
The generator $\Omega$ of ${\bf B}^{a}$ rotations, 
$[ \Omega, {\bf B}_{a} ] = \e_{ab} {\bf B}^{b}$ 
is obvious,
\be
\O = - \frac{1}{2} \; \int \; d^3x \, 
{\bf A}^{a} \cdot {\bf B}^{b} \delta_{ab} \;\; . 
\ee 
Equally obvious is the vanishing of its commutator with the 
$\frac{1}{2} \int \e_{ab} {\bf B}^{a} \cdot {\bf \dot A}^{b}$ 
kinetic term. Finally, as advertised, the invariance of the 
Hamiltonian density is a triviality: $\tilde{H}$ can only 
depend on the two manifestly duality invariant combinations 
$(u,v)$ of the three independent space scalars
\footnote{This can be obtained more ploddingly from
\[ [ \tilde{H}, \Omega ] = \e_{ab} {\bf B}^a \cdot 
\frac{\partial \tilde{H}}{\partial {\bf B}^b} = 0 \, , \] 
and solving the ensuing differential equation for $\tilde{H}$.}
\be                                                            
\tilde{H} = \tilde{H} (u,v) \;\; , \;\;\;\;
(t,u,v) \equiv \left( {\bf B}^2 \; , 
({\bf B}^a \cdot {\bf B}^a) \; ,
\frac{1}{4} (\e_{ab} {\bf B}^a \times {\bf B}^b)^2 \right) \;\; .
\ee

The Lorentz invariant $L$ depends on two 4--scalars $(\a,\b)$,
but of course neither necessarily implies the other. Indeed, 
the hard part is now to implement Lorentz invariance by 
(\ref{DiSc}). Note in
this connection that the momentum density, being
kinematical, is (like $\tilde{H}$) duality invariant, 
but is also independent of any assumed dynamics.

We find after some calculation that (\ref{DiSc}) 
constrains any $\tilde{H}(t,u,v)$ (dual or not) to obey 
\be
(\tilde{H}_u)^2 + u \tilde{H}_u \tilde{H}_v + 
v (\tilde{H}_v)^2 + \tilde{H}_t \tilde{H}_u + 
t \tilde{H}_t \tilde{H}_v = \frac{1}{4}
\label{Htuv}
\ee
which reduces in our case, $\tilde{H}(u,v)$ to
\be
(\tilde{H}_u)^2 + u \tilde{H}_u \tilde{H}_v + 
v (\tilde{H}_v)^2 = \frac{1}{4}   \; .
\label{Hdiff}
\ee

This result already follows from the weaker, ``half-integrated",
$[T^{00}({\bf r}), \tilde{H}] = \partial_{i} T^{0i}({\bf r})$
version of (\ref{DiSc}) which is of course the Hamiltonian
statement of the conservation requirement $\partial_{\m} T^{0\m}=0$.
(\ref{Htuv}) was also proposed in the present context, but 
from different considerations, some time ago in \cite{BB}.

Clearly the {\it Ans\"{a}tze} $\tilde{H} = \tilde{H}(u)$ 
and $\tilde{H} = \tilde{H}(u+v)$ yield the Maxwell and 
Born-Infeld solutions $\tilde{H}(u) = \frac{1}{2} u $, and 
$\tilde{H} =  \sqrt{1+u+v} -1$, respectively. [The overall 
$\tilde{H}$ normalization is forced by the kinetic terms.] 
To summarize, any $\tilde{H}$ depending only on $u$,$v$ and obeying 
(\ref{Hdiff}) defines a duality and Lorentz invariant model.

Going back to the full Lagrangian formulation, 
using our inputs (1--4) will yield the duality 
constraint in terms 
of the original second order $L(\a,\b)$ of (\ref{act}).  
The calculations are a bit tedious and we merely sketch 
the steps. Express $u, v$ in terms of 
$(t \equiv {\bf B}^2, \a , \b)$:
\begin{eqnarray}
v & = & 4L_{\alpha}^2 (t^2 - \alpha t - \beta^2) \;\; , 
\nonumber \\
u & = & t (1 + 4 L_{\alpha}^2 + L_{\beta}^2) - 
4 \alpha L_{\alpha}^2 - 4 \beta L_{\alpha} L_{\beta} \;\; .
\label{uv}
\end{eqnarray}
Next write $(d\a, d\b)$ in terms of $(dt, du, dv)$ 
by solving for these differentials using (\ref{uv}).
Using (\ref{ham}), namely $\tilde{H}(t,\a,\b) 
= - 2 L_{\alpha} t - L + 2 \alpha 
L_{\alpha} + \beta L_{\beta}$, rewrite $d \tilde{H}(t,\a,\b)$
as $d \tilde{H}(t,u,v) = \tilde{H}_t dt + \tilde{H}_u du 
+ \tilde{H}_v dv$. But in this basis, $\tilde{H}_t = 0$, 
while our calculation yields
\be
\tilde{H}_t =  
\frac{\b (1- 4 L_{\a}^2 + L_{\b}^2) + 2 \a L_{\a} L_{\b}} 
{2 (2 \b L_{\a} - t L_{\b})} \;\; .
\ee
Hence we have reproduced the covariant conditions 
of \cite{GR,PS}:
\be
L_{\alpha}^2 - \frac{\alpha}{2 \beta} L_{\alpha} 
L_{\beta} -
\frac{1}{4} L_{\beta}^2 = \frac{1}{4} \; .
\label{Ldiff}
\ee
[As a consistency check, we note that (\ref{Ldiff})
also implies $(\tilde{H}_u, \tilde{H}_v) 
= \left( \frac{\b}{2} \; , \frac{-L_{\b}}{4 L_{\a}} \right) 
\, (t L_{\beta} - 2 \beta L_{\alpha})^{-1}$;  
substituting these into (\ref{Hdiff}) 
shows that it is satisfied provided (\ref{Ldiff})
holds.] In this form, the Maxwell and Born-Infeld solutions are
$L = - \frac{1}{2} \a$ and $L = 1 - \sqrt{1+\a -\b^2}$, respectively.

An amusing parallel to our procedure arises for a massless
scalar field whose (3+1) variables are $(\pi,\phi)$, with 
canonical Lagrangian $L= \pi {\dot \phi} - 
H(\pi^2,({\nabla} \phi)^2)$. The Dirac covariance requirement
is the familiar \cite{PS} equation
\be
H_x H_y = \frac{1}{4} \;\;\;, \;\;
(x,y) \equiv (\pi^2,({\nabla} \phi)^2) \;\;\; .
\ee
Particular solutions include $H=\frac{1}{2}(x+y)$,
the free field (``Maxwell"), and $H=\sqrt{(1+x)(1+y)}-1$,
(``Born-Infeld"). Covariantly, 
$L=-\frac{1}{2}(\partial_{\m} \phi)^2$ and
$L=1- \sqrt{-det[\eta_{\m\n}+(\partial_{\m} \phi)(\partial_{\n} \phi)]}
=1-\sqrt{1+(\partial_{\m} \phi)^2}$, respectively, as follows from
the Legendre transform equivalent of (\ref{firact}): From a general
$L(z)$, $z \equiv \frac{1}{2}(\partial_{\m} \phi)^2$, 
\be
I[\pi^{\m},\phi] = \int d^4x \tilde{L} = \int d^4x \left(
L' \pi^{\m} (\partial_{\m} \phi) + L -2zL' \right) \;,
\ee
where $\pi^{\m}$ and $\phi$ are to be varied independently,
$L$ is to be regarded as a function of $z$ only,
$'$ denotes differentiation with respect to $z$, and the
$\pi_{\m}=\partial_{\m} \phi$ relation follows from 
varying $\pi_{\m}$.

Although we have explicitly worked with one--form 
potentials in $D=4$, the same procedure can be applied 
to obtain the duality criteria 
for actions with $(2n+1)$--form potentials in $D=4(n+1)$ 
spaces, using a spacetime decomposition with 
electric/magnetic $(2n+1)$ forms.
Unfortunately, the number of spacetime as well as 
spatial invariants grow so rapidly with dimension that
the explicit steps become untractable beyond $D=4$.

We thank A. Gomberoff, M. Henneaux, J. McCarthy, and 
C. Teitelboim for useful discussions, and NSF for 
support under grant \#PHY-9315811.

\end{document}